\documentstyle[aps,prb,preprint,epsf]{revtex}

\begin{document}


\author{A. G. Rojo}
\title{Absence of gap for infinite half--integer spin ladders with an
odd number of legs}
\address{Department of Physics, The University of
Michigan,
Ann Arbor, MI 48109-1120
}

\maketitle
\date{The Date }
\maketitle

\begin{abstract}
A  proof is presented for the absence of gap for spin $1/2$ ladders
with an odd number of legs, in the infinite leg length limit.
This result is relevant to  the current discussion of coupled
one--dimensional spin systems, a physical realization of which are
vanadyl pyrophosphate, (VO)$_2$P$_2$O$_7$, and stoichiometric
 Sr$_{n-1}$ Cu$_{n+1}$ O$_{2n}$
(with $n=3,5,7,9,\dots$).
\end{abstract}

\vskip2pc \narrowtext
The magnetic properties of low dimensional systems have
been  the subject of  intense  theoretical and experimental
research in recent years.
Haldane's conjecture\cite{haldane}, which states
that  one--dimensional chains of integer spin should have
a gap in the excitation spectrum, whereas the half integer spin  chains
should be gapless, was confirmed by a number of experiments in
 quasi one--dimensional systems\cite{exp1}.
The recent  interest on  the spin $1/2$  Heisenberg model in two dimensions
(2D)
is largely due to the fact that the undoped parent compounds of
the high-Tc superconductors are accepted to be described by this
model\cite{anderson,manousakis}. Although an exact solution
is lacking in 2D, numerical results\cite{manousakis} indicate
long range order in the ground sate, and  gapless excitations.
Most of these calculations involve  scaling  the results obtained for an
$M\times M$ square lattice to the limit $M\rightarrow \infty$.

More recently, the crossover from one to two dimensions has been
studied in ladder type geometries of $N_x \times N_y $ sites, by taking
the limit of $N_x\rightarrow \infty$ first\cite{dagotto1,white}.
Remarkably, these results show that the scaling as a function of
$N_y$ is not smooth: ladders of even $N_y$ have a finite gap,
while those
of odd $N_y$ have zero gap\cite{dagotto2}.
This interesting property turns out to be more than an
academic curiosity, since, as pointed out in Ref. \cite{rice1},
real compounds like stoichiometric Sr$_{n-1}$Cu$_{n+1}$O$_{2n}$
(with $n=3,5,7,9,\dots$)
can be described by Heisenberg spin ladders with $N_y=\frac 12 (n+1)$.
Also, (VO)$_2$P$_2$O$_7$ (vandyl pyrophosphate), has a ladder ($N_y~=~2$)
configuration
of spin~$1/2$, V$^{+4}$ ions\cite{johnston}.

A qualitative explanation of  the difference between even and odd
$N_y$ resorts to the strong coupling limit $J'/J \gg 1$, with
$J'$ being the exchange integral in the direction of the
rungs of the ladder\cite{dagotto2,reigro}.
 In the limit $J\rightarrow 0$ the
 system decouples into open  chains of $N_y$ sites. Chains with odd
$N_y$  are gapless due to Kramers degeneracy,
whereas those of even $N_y$ have a gap; for example, for $N_y=2$, the
spin gap $E_g=J'$.
For odd $N_y$ and  $J'/J \gg 1$, the Kramers doublets in different chains can
be coupled perturbatively, the resulting effective Hamiltonian being a
one--dimensional infinite chain of spin~$1/2$ that has zero gap.
Since for both the opposite extremes
of $J'/J \gg 1$ and $J'=0$ the system is gapless for odd $N_y$,
one expects a gapless spectrum for intermediate values of $J'$.
An alternative explanation is offered in Ref. \cite{dima}
using arguments similar to those of Haldane for  one dimensional
chains: for odd $N_y$ chains, the topological
term
governing the long-wavelength dynamic of the lattice
does not vanish for odd $N_y$ and the system is gapless. For even
$N_y$ a gap remains that scales as $E_g(N_y) \sim \exp{(-N_y)}$.
In this paper, I show that the absence of a gap for
odd $N_y$  is a rigorous result.
This I do by
reanalizing  the
Lieb,
Shultz and Mattis (LSM) theorem\cite{lieb}~---~which is valid
for one dimensional spin-$1/2$ chains---for
 ladder
geometries with $N_y=2P+1$ sites in the $y$--direction, and $N$ sites in the $%
x$--direction\cite{lsm}, in the limit $N\rightarrow \infty$.
The proof is valid if $N$ is even. For odd $N$
the total number of sites is odd,  the ground state
has total spin projection  $S_{{\rm Tot}}^z=\pm 1/2$, and is
therefore doubly degenerate\cite{lm}.
%

We consider the celebrated Heisenberg model,  which is described by the
Hamiltonian%
\begin{equation}
H=J\sum_{\langle {\bf i},{\bf j}\rangle }S({\bf i)}\cdot S({\bf j)}
,
\label{h1}
\end{equation}
where ${\bf i}$ and ${\bf j}$ are vectors of the lattice, $S({\bf i})$
denotes spin--$1/2$ operators, and the symbol $\langle {\bf i},{\bf j}%
\rangle $ stands for  near neighbors. The boundary conditions are periodic,
and  the lattice constant equals to $1$, in such way that ${\bf i=(}n_x,n_y)$%
, with $n_x$ and $n_y$ integers.

Let $|\Psi _0\rangle $ be the ground state of $H$ with energy $E_0$. Define
a state $|\Psi _x(k)\rangle $ which is obtained from   $|\Psi _0\rangle $ by
applying a twist of wave vector $k$  in the $x$--direction:%
\begin{equation}
|\Psi _x(k)\rangle =\exp [ik\sum_{{\bf i}}n_xS^z({\bf i)]}|\Psi _0\rangle
\equiv U_x(k)|\Psi _0\rangle .
\end{equation}
Note that spins
with the same $n_x$ coordinate are subject to the same twist.

The proof of existence of gapless excitations for $N\rightarrow \infty $ can
be divided in two steps. In step one, the orthogonality of $|\Psi
_x(k)\rangle $ with $|\Psi _0\rangle $ is established. It is precisely in
this step where the distinction between even and odd widths becomes
apparent: for even widths, the twisted state  $|\Psi _x(k)\rangle $ will not
be
orthogonal to the ground state.  In step two,  the expectation value of $H$ in
the twisted wave function is proven to be equal to $E_0$ in the limit $%
N\rightarrow \infty $.

{\em Step 1. }Consider the operator $T_x$, which translates the system by a
lattice constant in the $x$--direction:

$$
\begin{array}{c}
T_xS(n_x,n_y)T_x^{-1}=S(n_x+1,n_y), \\
T_xS(N,n_y)T_x^{-1}=S(1,n_y).
\end{array}
$$

Since $T_x$ commutes with $H$, and
we have $T_x$ $|\Psi _0\rangle =e^{i\delta }$
$|\Psi _0\rangle $, with $\delta $ some constant phase.
Here we are using the assumption of uniqueness of the ground state, which
can actually be proven for finite $N$ (see below).
This allows us to
write%
\begin{equation}
\langle \Psi _0|\Psi _x(k)\rangle =\langle \Psi _0|T_x\exp [ik\sum_{{\bf i}%
}n_xS^z({\bf i)]}T_x^{-1}|\Psi _0\rangle .
\end{equation}

The evaluation of $T_x\exp [ik\sum_{{\bf i}}n_xS^z({\bf i)]}T_x^{-1}$ is
straightforward, giving%
\begin{equation}
T_xU_x(k)T_x^{-1}=U_x(k)\exp [-ik\sum_{{\bf i}}S^z({\bf i)]\exp [}%
iNk\sum_{n_y=1}^{N_y}S^z(1,n_y)].
\end{equation}
Since, by Marshall's theorem, the ground state is in the subspace of $S^z_{%
{\rm Tot}}=0$, we have%
\begin{equation}
\exp [-ik\sum_{{\bf i}}S^z({\bf i)]}|\Psi _0\rangle =|\Psi _0\rangle .
\end{equation}
We now prove that, if one chooses $k=2\pi m/N$, with $m$ an odd integer, then%
\begin{equation}
{\bf \exp [}iNk\sum_{n_y=1}^{N_y}S^z(1,n_y)]|\Psi _0\rangle =-|\Psi
_0\rangle .
\end{equation}
Let%
$$
|\Psi _0\rangle =\sum_\Gamma c(\Gamma )|\Gamma \rangle ,
\label{psi}
$$
where the sum runs over all the spin configurations $\Gamma $
with $S^z_{\rm Tot}=0 .$ Then%
\begin{equation}
{\bf \exp [}iNk\sum_{n_y=1}^{N_y}S^z(1,n_y)]|\Psi _0\rangle =\sum_\Gamma
\exp [i2\pi mQ_\Gamma ]c(\Gamma )|\Gamma \rangle ,
\label{gama}
\end{equation}
with $Q_\Gamma =\frac 12(N_{\uparrow ,\Gamma }-N_{\downarrow ,\Gamma })$,
and $N_{\uparrow ,\Gamma }$ ( $N_{\downarrow ,\Gamma }$ ) the number of up
(down) spins in row $1$ in configuration $\Gamma .$ If $N_y(\equiv $ $%
N_{\uparrow ,\Gamma }+N_{\downarrow ,\Gamma })$ is {\em odd}, then $Q_\Gamma
$ is half integer, and $\exp [i2\pi mQ_\Gamma ]=-1.$ This implies that
\begin{equation}
\langle \Psi _0|\Psi _x(k)\rangle =-\langle \Psi _0|\Psi _x(k)\rangle =0 \,
,
\end{equation}
and the states are in fact orthogonal. We stress that the proof in this step
is valid only for {\em odd }widths.
For {\it even} widths  $Q_\Gamma$ is  integer  and
$\exp [i2\pi mQ_\Gamma ]=1$. We could have chosen $k=\pi m /N$
in this case, but still the sign of $\exp [i2\pi mQ_\Gamma ]$
will depend on $\Gamma $ through  $N_{\downarrow ,\Gamma }$, and cannot be
taken out
of the summation in (\ref{gama}).
The failure of the present  proof for even widths has
certainly the same  formal origin than the case of
strictly one dimensional chains of
integer
spin.

{\em Step 2. }
We now need to evaluate $U_x(k)^{-1}HU_x(k).$ The ``$S^zS^z$''
component of $H$ remains unchanged
since  $U_x(k)$ commutes with $S^z({\bf i)}$. The transformation of the $xy$
component is
easy to obtain by noting that
\begin{equation}
e^{-i\alpha S^z({\bf i)}}S^{\pm }({\bf i)}e^{i\alpha S^z({\bf i)}}=e^{\mp
i\alpha
}S^{\pm }({\bf i),}
\end{equation}
for $\alpha $ an arbitrary constant.
This implies that the $xy$ component of $H$ corresponding to the rungs of
the ladder is also unchanged.%
 For the transformed Hamiltonian we have
$$
\begin{array}{c}
U_x(k)^{-1}HU_x(k)=H
{\bf +}\frac 12 J[\cos
(k)-1]\sum_{n_x=1}^N\sum_{n_y=1}^{N_y}[S^{+}(n_x,n_y)S^{-}(n_x+1,n_y)+{\rm %
h.c}]{\bf +} \\ \frac i2 J\sin
(k)[\sum_{n_x=1}^N\sum_{n_y=1}^{N_y}[S^{+}(n_x,n_y)S^{-}(n_x+1,n_y)-{\rm h.c}%
].
\end{array}
$$

Since the coefficients $c(\Gamma )$ are real,
\begin{equation}
\langle \Psi _0|[S^{+}(n_x,n_y)S^{-}(n_x+1,n_y)-{\rm h.c}]|\Psi _0\rangle
=0.
\end{equation}
Now we expand the cosine for small $k$, in particular we will take $m=1$  ($%
k=2\pi /N$): $\cos (k)-1\approx 2\pi ^2/N^2.$ Also, we make use of the
inequality
\begin{equation}
\frac 12\langle \Psi _0|[S^{+}(n_x,n_y)S^{-}(n _x+1,n_y)+{\rm h.c}]|\Psi
_0\rangle \leq S^2
,
\end{equation}
to write a bound for the energy of the twisted state:%
\begin{equation}
\langle \Psi _x(k)|H|\Psi _x(k)\rangle =\langle \Psi
_0|U_x(k)^{-1}HU_x(k)|\Psi _0\rangle \leq E_0+\frac{2\pi ^2JS^2N_y}%
N+O(N_y/N^3).
\end{equation}
It is clear that if the limit $N\rightarrow \infty $ is taken for fixed $N_y$,
the twisted state is degenerate with the ground state, and the proof is
complete for the existence of gapless excitations for infinite ladders with
an odd number of legs. Note that the proof can be extended immediately to
the following cases:
(a) coupling constants that are different in the vertical and
horizontal directions, as long as
translational invariance in the $x$--direction
is preserved, (b) half integer spins higher than $\frac 12$,
(c) quasi one dimensional ``wires" of a two dimensional cross section
with an odd number of spins.

The question of the uniqueness of the ground state was addressed for the
case of chains by Affleck and Lieb\cite{al} (AL), whose considerations apply
also
to ladders.
In particular, one can prove that the ground state is unique for {\it finite}
$N$. The proof is as follows. For $N$ even we can define two sublattices $A$
(for which $n_x+n_y$ is even) and
$B$
(for which $n_x+n_y$ is odd),
 and rotate all the spins in the $A$ sublattice in such a way that
$S^z({\bf i}) \rightarrow S^z({\bf i})$, and $S^{\pm}({\bf i})
 \rightarrow -S^{\pm}({\bf i})$.
 This canonical transformation has the effect of changing the sign of the
 ``$S^+S^-$" term in (\ref{h1}).
 After this transformation, all the off diagonal elements between
configurations
 $\Gamma$ and $\Gamma'$ which are connected by the $S^+S^-$ term are negative.
 Therefore, in the minimum energy state, all the weights $c(\Gamma)$ will have
 the same sign. If we call $E(\Gamma)$ the diagonal energy of
 configuration $\Gamma $, and $\Gamma '$ a configuration obtained by
 applying the $S^+S^-$ term to  $\Gamma $, we have
 that $c(\Gamma)$ obeys the equation of motion\cite{lsm}
 $\left[ E(\Gamma)-E_0\right]c(\Gamma)=J/2\sum_{\Gamma '}c(\Gamma')$.
 Since all $C(\Gamma)$ have the same sign, this implies that all
 $c(\Gamma)\neq 0$. Therefore the ground state is non--degenerate.
 However, it could happen that, as $N\rightarrow \infty$ , some weights
$c(\Gamma)$
 become exponentially small
 and the ground state splits up into two broken symmetry states.
 Possibilities are an antiferromagnetically  ordered state, which will
 have gapless spin-wave excitations, and a spontaneously  dimerized one, which
 which will have a gap. We have therefore proven that in the limit
 $N\rightarrow \infty$ either there are degenerate ground states or
 vanishing gap.

I acknowledge helpfull conversations with E. Dagotto.


\begin{references}
\bibitem{lsm}
  In Appendix B of Ref\cite{lieb} it is suggested that systems of $N$ sites in
one
  direction and $N'~=~O(N^\nu)$ sites in the other direction, where $0<\nu<1$,
  will be gapless if $N\rightarrow \infty$. However, the crucial distinction
between even and odd
  widths is missed.
\end{references}
\end{document}